\documentclass[12pt]{article}
\title{Photon-Neutrino Interactions in Magnetic Field through Neutrino 
	Magnetic Moment}
\author{Ashok Goyal\thanks{E--mail : agoyal@ducos.ernet.in}\\ 
	{\em Department of Physics and Astrophysics,} \\
	{\em University of Delhi, Delhi-110 007, India.} \\
        {\em InterUniversity Centre for Astronomy and Astrophysics,} \\
        {\em Ganeshkhind, Pune 411007 , India.} \\
        }

\begin{document}
\maketitle
\large


\begin{abstract}
\noindent
We study the neutrino-photon processes like $\gamma\gamma\rightarrow\nu\bar{\nu}$ in the presence of uniform external magnetic field for the case when neutrinos can couple to the electromagnetic field directly through their dipole magnetic moment and obtain the stellar energy loss. The process would be of special relevance in astrophysical situations where standard left-handed neutrinos are trapped and the right handed neutrinos produced through the spin flip interaction induced by neutrino magnetic moment alone can freely stream out.
\end{abstract}
\pagebreak

Photon-neutrino interactions are of interest in the astrophysical and cosmological environment. Neutrino pair production through $\gamma\gamma\rightarrow\nu\bar{\nu}$ would provide energy loss mechanism in stellar bodies and its reverse process $\nu\bar{\nu}\rightarrow\gamma\gamma$ and the neutrino photon scattering $\gamma\nu\rightarrow\gamma\nu$ would be  important in stellar evolution. The scattering process may also be important for studying photon and neutrino propagation over intergalactic distances. The amplitude for any of these processes is however, known to be highly suppressed [1-3]  because of the 
vector-axial vector nature of weak interactions and the fact that the photons cannot
 couple to J=1 state (Yang's theorem [4]). In the four fermion limit of the 
weak interactions, the amplitude is exactly zero for massless neutrinos [5] 
and is suppressed by an additional factor of $\frac{1}{(M_w)^2}$ in the Standard model [1-3,6-8]. The resulting
cross-sections are therefore exceedingly small and are unlikely to be of any importance
in astrophysics. The process $\gamma\gamma\rightarrow\nu\bar{\nu}$, when neutrinos are
 massive or when they have more general interactions, has also been discussed in the
 litrature [9]  but there does not seem to be any scenario where this process
 can be of astrophysical interest. Three photon coupling to neutrinos however, does not
 suffer from these suppressions and it has been shown recently by Dicus and Repko [10] that inelastic processes like $\gamma\gamma\rightarrow\nu\bar{\nu}\gamma$
 , $\gamma\nu\rightarrow\gamma\nu\gamma$ etc. are much larger than the corresponding 
elastic processes for photon energy greater than 1 KeV. It is well known that in many 
astrophysical environments photon-neutrino reactions along with other weak processes 
relevant in astrophysics take place in the presence of strong magnetic field. Magnetic 
fields of strength $B\sim B_c=\frac{(m_e)^2}{e} = 4.41\times {10}^{13} G$ or more are 
known to exist in compact stars and many of the processes have been studied in the
 presence of magnetic field [11].

In the presence of magnetic field, $\gamma\gamma\rightarrow\nu\bar{\nu}$ elastic processes 
are akin to $\gamma\gamma\rightarrow\nu\bar{\nu}\gamma$ inelastic processes with one of the photon legs being replaced by the external magnetic field and are
therefore not expected to be supressed. Starting from the Euler-Heisenberg effective Lagrangian [12] for photon-photon scattering and by replacing one photon polarisation by the neutral current, Dicus and Repko [10] obtained an effective action for $\gamma\gamma\rightarrow\gamma\nu\bar\nu$ inelastic processes and calculated the cross-sections for three photon inelastic processes. Subsequently Shaisulanov [13] calculated the two photon elastic cross-sections in the presence of external uniform magnetic field and showed that the cross-sections are enhanced by a factor $\sim (\frac{M_w}{m_e})^4.(\frac{B}{B_c})^2$ compared to vacuum and are comparable with the three photon inelastic cross-sections.\\
\indent The presence of electro-magnetic dipole moment implies that the neutrinos can couple directly to electromagnetic field allowing for a variety of nonstandard processes. One of the most interesting process in this context is the plasmon decay $''\gamma ``\rightarrow\nu\bar\nu$ which happens to be the dominant neutrino emission process from stars for a wide range of temperatures and densities. Many of these processes have been studied in detail in the litrature and bounds on neutrino magnetic moment have been obtained from consideration of stellar energy loss. For details see [14]. In this context it is interesting to investigate the photon-neutrino elastic processes through direct neutrino coupling to photons via neutrino magnetic moment. The leading diagram in QED involving three photons attached to fermion loop vanishes because of Furry's theorem. The next non-vanishing diagram involves four photon vertex through the fermion box , with one of the photons replaced by external magnetic field. Following [10,13] we can calculate $\gamma\gamma\rightarrow\nu\bar\nu$ cross-section in the presence of external magnetic field by using the Euler-Heisenberg effective Lagrangian for photon-photon scattering, namely \\
\begin{eqnarray}
\mathcal L_{eff}^{\gamma\gamma} = \frac{\alpha^2}{180m_e^4}[5f^{\mu\nu}f_{\mu\nu}f^{\lambda\rho}f_{\lambda\rho}-14f^{\mu\nu}f_{\nu\lambda}f^{\lambda\rho}f_{\rho\mu}]
\end{eqnarray}
and the magnetic moment interaction of the neutrino with electromagnetic field given by
\begin{eqnarray}
\mathcal L_{eff}^{\gamma\nu} = \mu_{\nu}\bar\psi\sigma^{\mu\nu}\psi f_{\mu\nu}
\end{eqnarray}
Now replacing one of the photon field tensors $f_{\mu\nu}$ in (1) by the external magnetic field tensor $F_{\mu\nu}$ and attaching the neutrino tensor current in (2) to another photon field tensor and taking all distinct permutations, we get as in [13] 
\begin{eqnarray}
\mathcal L_{eff} = \frac{4\mu_{\nu}\alpha^2}{180m_e^4}.N^{\mu\nu}[5(F_{\mu\nu}f_{\lambda\rho}f^{\lambda\rho} + 2f_{\mu\nu}f^{\lambda\sigma}F_{\lambda\rho})\nonumber \\
 -14 (F_{\nu\lambda}f^{\lambda\rho}f_{\rho\mu} + F^{\lambda\rho}f_{\nu\lambda}f^{\rho\mu} + F_{\rho\mu}f^{\lambda\rho}f_{\nu\lambda})].\frac{1}{q^2-m_{\gamma}^2}
\end{eqnarray}
where q is the four momentum carried by the neutrino pair, $m_{\gamma}$  is the plasmon mass which depends on the state of the plasma and $\mu_{\nu}$ is the neutrino magnetic dipole moment The neutrino tensor $N^{\mu\nu}$ is given by
\begin{eqnarray}
N^{\mu\nu}= [\partial^{\mu}(\bar{\psi} \sigma^{\nu\alpha}q_{\alpha}\psi) - \partial^{\nu}(\bar{\psi} \sigma^{\mu\alpha}q_{\alpha}\psi)]
\nonumber\\
=[(p_{1}+p_{2})^{\mu}(p_{1}-p_{2})^{\nu}-(p_{1}+p_{2})^{\nu}
(p_{1}-p_{2})^{\mu}]\bar{u}(p_{1})v(p_{2})
\end{eqnarray}
The amplitude for the process $\gamma\gamma\rightarrow\nu\bar\nu$ can be written down  by using (3) and (4) and the cross-section averaged over the polarisations of the 
incoming photons can be evaluated in a straight forward way. After a long and involved calculations we get
\begin{eqnarray}
\sigma(\gamma\gamma\rightarrow\nu\bar{\nu}) &=& \frac{8\mu_{\nu}^{2}\alpha^{4}}
{3\pi} \frac{1}{(180 m_{e}^{4})^2} \frac{(k_{1}.k_{2})^2}{(2k_{1}.k_{2}
-m_{\gamma}^2)}
\nonumber \\
&& [k_{1}.k_{2} [6416 ( k_{1} F^{2} k_{1} + k_{2} F^{2} k_{2} ) 
+ 6272 k_{1} F^{2} k_{2}]
\nonumber  \\
&&-3424 (k_{1} F k_{2})^2+784 (k_{1}.k_{2})^2
F_{\mu\nu}F_{\mu\nu}]
\end{eqnarray}

The cross-section can be estimated in the c.m.frame and by considering the magnetic field in the z-direction say
\begin{eqnarray}
\sigma(\gamma\gamma\rightarrow\nu\bar\nu) 
 &&=\frac{2\mu_\nu^2\alpha^4}{3\pi}\frac{1}{(180m_e^4)^2}\frac{1}{(1-\frac{m_\gamma^2}{4w^2})^2} 288w^4B^2 \nonumber \\
 &\sim& 0.31 \times 10^{-50} \mu_{10}^2(\frac{w}{m_e})^4(\frac{B}{B_c})^2.\frac{1}{(1-\frac{m_\gamma^2}{4w^2})^2} cm^2
\end{eqnarray}
where $\mu_{\nu}=\frac {\mu_{10}}{10^{-10}}\mu_{B}$  with $\mu_{B} = 
\frac{e}{2m_{e}}$ and $w$ is the photon energy.
The energy loss from a magnetised star can now be obtained in a straight forward manner by using the expression
\begin{eqnarray}
\mathcal Q = \frac{1}{(2\pi)^6}\int\frac{2d^3k_1}{e^{w_1/T}-1}\int\frac{2d^3k_2}{e^{w_2/T}-1}.\frac{k_1.k_2}{w_1w_2}(w_1+w_2)\sigma(\gamma\gamma\rightarrow\nu\bar\nu)
\end{eqnarray}
In the range of validity  of the effective action approach used above the plasmon effects are negligible and can be dropped. In this approximation we get
\begin{eqnarray}
\mathcal Q 
&&=\frac{512\pi}{1148175}\mu_\nu^2\alpha^4\frac{B^2}{M_e^8}T^{11}[25263\zeta(7)-1576\pi^2\zeta(5)]\nonumber \\
&\sim&  0.32 \times 10^{10}\mu_{10}^2(\frac{B}{B_c})^2 T_9^{11} ergs s^{-1} cm^{-3}
\end{eqnarray}
where $T_9$ is the temperature in units of $10^9$ K.
\indent Direct laboratory bounds on neutrino magnetic moment give [15]
$\mu_{\nu}\leq 1.8\times 10^{-10}\mu_B$ , $ \mu_{\nu}\leq 7.4\times 10^{-10}\mu_B$ , $ \mu_{\nu}\leq 5.4\times 10^7\mu_B$ for electron, muon and tau neutrinos respectively. In addition to the direct laboratory limits given above, there exists limits based on astrophysical and cosmological considerations arising from stellar cooling and nucleosynthesis arguments. From SN1987A and red giants in globulor clusters we obtain typically [14,15], $\mu_{\nu}\leq (2-.3)10^{-12}\mu_B$ and $\mu_{\nu}\leq 3\times 10^{-11}\mu_B$ . However,these limits are to varying degree model dependent and apply to all neutrino flavors. Thus for allowed neutrino magnetic moment values, the two photon rates in the presence of strong magnetic field are roughly of the same order as the enelastic processes discussed in [10] and could be important in astrophysics. In the enviornment of a proto-neutron star where temperatures could be $\geq 10^{11}$ K and the standard model left handed neutrinos are trapped in the star due to their small mean free path, production of right handed neutrinos through spin-flip electromagnetic interactions induced by neutrino magnetic moment would result in the rapid cooling of neutron stars due to the emission of right handed neutrinos. The cross-section and the cooling rate relevant at these temperatures can not be reliably estimated from the effective action considered here. Recently [15] two photon elastic processes in the standard model have been studied in the presence of background magnetic field in the kinematic regime relevant in stars with temperatures much above $m_e$ and it has been shown that for temperatures $\sim 10^{11}$ K, the effective action approach overestimates the rates by several orders of magnitude. Clearly a more careful analysis is required to study the photon-neutrino interaction processes in the magnetised supernova environment and is being undertaken.
\begin{section}*{Acknowledgements}
We would like to thank Professor J.V. Narlikar for providing hospitality at
the Inter-University Centre for Astronomy and Astrophysics, Pune 411
007, India where this work was completed. 
\end{section}
\pagebreak


\begin{thebibliography}{99}
\bibitem{1}  M.~J.~Levine, Nuovo Cimento $\mathbf{29A}$, 67 (1967).
\bibitem{2}  V.~K.~Cung and M.~Yoshimura, Nuovo Cimento $\mathbf{29A}$, 557 (1975).
\bibitem{3}  J.~Liu,  Phys. Rev. $\mathbf{D44}$, 2879 (1991).
\bibitem{4}  C.~N.~Yang, Phys. Rev. $\mathbf{77}$, 242 (1950) ; L.~D.~Landau, Sov. Phys. Dokl. $\mathbf{60}$, 207 (1948).
\bibitem{5}  M.~GellMann, Phys. Rev. Lett. $\mathbf{6}$, 70 (1961).
\bibitem{6}  J.~F.~Nieves, P.~Pal and D.~G.~Unger, Phys. Rev. $\mathbf{D28}$, 908 (1983).
\bibitem{7}  P.~Langacker and S.~Liu, Phys. Rev. $\mathbf{D46}$, 4140 (1992).
\bibitem{8}  D.~A.~Dicus and W.~W.~Repko, Phys. Rev. $\mathbf{D48}$, 5106 (1993).
\bibitem{9}  A.~Halprin, Phys. Rev. $\mathbf{D11}$, 147 (1975); E.~Fischbach et al, Phys. Rev. $\mathbf{D13}$, 1523 (1976); ibid $\mathbf{16}$, 2377 (1977); A.~A.~Natale, V.~Pleitez and A.~Tacla, Phys. Rev. $\mathbf{D36}$, 3278 (1987).
\bibitem{10} D.~A.~Dicus and W.~W.~Repko, Phys. Rev. Lett. $\mathbf{79}$, 569 (1997).
\bibitem{11} S.~L.~Adler, Ann. Phys. $\mathbf{67}$, 599 (1971); V.~N.~Baier, A.~Milstein and R.~Zh.~Shaisultanov, Phys. Rev. Lett. $\mathbf{77}$, 1691 (1996); S.~L.~Adler and  ~C.~Schubert, Phys. Rev. Lett. $\mathbf{77}$, 1695 (1996);
~A.~N.~Ioannisian and ~G.~G.~Raffelt, Phy. Rev. D$\mathbf{55}$, 7038 (1997);
~A.~Vilenkin, Astrophys. J.$\mathbf{451}$, 700 (1995); ~V.~G.~Bezchastnov
and ~P.~Haensel, Phys. Rev. D$\mathbf{54}$, 3706 (1996);
~C.~J.~Horowitz and ~J.~Piekavewicz, Nucl. Phys. B$\mathbf{640}$, 281 (1998);
~A.~Goyal, Phys. Rev. D$\mathbf{59}$, 101301 (1999).
\bibitem{12} H.~Euler, Ann. Phys. $\mathbf{26}$, 398 (1936); W.~Heisenberg and H.~Euler, Zeit. Phys. $\mathbf{98}$, 714 (1936).
\bibitem{13}  R.~Shaisultanov, Phys. Rev. Lett. $\mathbf{80}$, 1586 (1998).
\bibitem{14}  See for example G.~G.~Raffelt,$\it {Stars~as~Laboratories~for
~Fundamental~Physics}$ (University of Chicago Press, Chicago, 1996) and references therein.
\bibitem{15}  Tzuu-Kang Chyi etal, Phys. Lett. $\mathbf{B466}$, 274 (1999). 
\bibitem{16}  See Reviews of Particle Physics, The Euro. Phys. Jour. C
$\mathbf {3}$, 1 (1998).
\end{thebibliography}
\end{document}